\documentclass[reprint,superscriptaddress,nofootinbib,aps,pra]{revtex4-2}
\pdfoutput=1

\usepackage{graphicx, placeins, float}
\usepackage{amssymb,amsmath}
\graphicspath{ {./images/} }

\usepackage{xcolor}
\usepackage{hyperref}
\hypersetup{
	colorlinks = true,
	linkbordercolor = {white},
	linkcolor={black},
	citecolor={black},
	urlcolor={black}
}

\begin{document}
\title{Quantum Measurements of Time: A reply to criticisms}
\author{Lorenzo Maccone}
	\affiliation{Dipartimento di Fisica and INFN Sezione di Pavia, Università degli Studi di Pavia, Via Agostino Bassi 6, I-27100, Pavia, Italy}
	
\author{Simone Roncallo}
	\affiliation{Dipartimento di Fisica and INFN Sezione di Pavia, Università degli Studi di Pavia, Via Agostino Bassi 6, I-27100, Pavia, Italy}
	
\author{Krzysztof Sacha}
	\affiliation{Instytut Fizyki imienia Mariana Smoluchowskiego, Uniwersytet Jagiello\'{n}ski, ulica Profesora Stanisława Łojasiewicza 11, PL-30-348 Krak\'{o}w, Poland}

\begin{abstract}
  In \cite{comment} Cavendish et al. raise three criticisms against
  our time of arrival proposal \cite{qtoa}. Here we show that all
  three criticisms are without merit. One of them is founded on a
  logical mistake.
\end{abstract}

\maketitle  
In \cite{qtoa} we advanced a proposal for describing the time of arrival (TOA) of a particle at some spatial position, e.g.~\cite{mielnik}. Since time is not an observable in textbook quantum mechanics, a straightforward quantum description of TOA measurements cannot be given. In the literature, TOA proposals relate such measurements to other observables, or construct suitable POVMs. In \cite{qtoa}, instead, we suggest that a quantum TOA observable {\it can} be obtained through the (slight) extension of quantum mechanics proposed by Page and Wootters \cite{pw,qtime}. In the following and in our papers, we use a natural TOA definition: we say that ``a particle
has arrived at the detector'' iff (by definition) ``it is at the detector's position $D$''.

In \cite{comment}, Cavendish et al. criticized our proposal with, essentially, three arguments:
\begin{enumerate}
\item The ``dramatically different'' criticism: Cavendish et al. point out that our predictions are ``dramatically different from those of other well known proposals in the literature'', and they show a regime where these differences are very evident.
\item The ``empirically implausible'' criticism: Cavendish et al. suggest that our proposal is empirically implausible because of its dependence on the total time of the experiment $T$, and they suggest that this is not to be expected in a TOA.
\item The ``non-arrival probability'' criticism: Cavendish et al. claim that our proposal would predict that the probability that a particle does not arrive tends to one for large total time $T$, which is not what one would expect if the particle has arrived at some point in time.
\end{enumerate}

In the following, we carefully reply to these criticisms:
\begin{enumerate}
\item We point out that ``the dramatically different'' criticism is hardly a criticism per se: in the absence of experimental evidence, the argument that a radical new proposal matches the previous ones (or not), cannot be used to support (or disprove) its validity. We also point out that Cavendish et al.'s analysis on these differences has already appeared in \cite{toacomparison}, which was entirely and solely devoted to the comparison among different TOA proposals.
\item Using very simple examples, we show that, far from being ``empirically implausible'', the dependence of the TOA related quantities on $T$ is to be expected. We suspect that Cavendish et al.'s claim to the contrary might arise from a definition of TOA that does not match ours (Ours: ``the particle has arrived iff it is at the detector's position'').
\item With a simple counterexample, we show that the ``non-arrival probability'' criticism is incorrect: Cavendish et al. erroneously equate the ``probability that the particle does not arrive'' to the ``probability that the particle is not found at the detector'', instead of looking at the ``probability that the particle is not found at the detector AND that it was never there''. A crucial logical mistake that invalidates their argument.
\end{enumerate}

\section*{The ``dramatically different'' criticism}
The fact that the results of \cite{qtoa} are ``dramatically different'' from previous proposals is a feature, not a flaw, of our method. It is a feature, because it makes it very easy to discriminate experimentally other TOA distributions. We had pointed this out ourselves in \cite{toacomparison}, where a careful comparison with previous TOA literature was detailed. Cavendish et al.'s results are a specific instance of \cite{toacomparison}, where very similar graphs appear with different, and more realistic, parameters.

Since our proposal \cite{qtoa} is based on the Page and Wootters mechanism \cite{pw,qtime}, it is an extension of quantum mechanics. Then, the question of whether it is correct or not can only be assessed experimentally: not through textbook quantum mechanics, and certainly not by questioning its agreement with previous untested
proposals.

In essence, we agree with Cavendish et al.'s claim that our predictions are ``dramatically different'' from previous ones (we devoted an entire paper, \cite{toacomparison}, to pointing that out). However, we are baffled on why this should be construed as a criticism to our proposal.

\section*{``Empirically implausible'' criticism}
We agree with Cavendish et al. that, in some regimes, the results of our proposal \cite{qtoa} depend explicitly on the total duration $T$ of the experiment.  In that paper, we had already pointed out this
fact, and explicitly analyzed these regimes (e.g. end of pg.2 and beginning of pg.3 of \cite{qtoa}).  This is an expected characteristic of our proposal, which directly stems from our above definition of
time of arrival: ``a particle has arrived iff it is at the detector position $D$''. The probability of finding the particle at $D$ certainly depends on how long one is looking for it, even in the
classical regime which is simpler to visualize: Suppose that the particle spends one second at the detector position. If the total duration $T$ of the experiment is exactly the same one second, the probability of finding the particle at the detector is $100\%$: the particle is already there. If instead, one considers a total duration time of 10 seconds, the probability of finding the particle at the detector is 1/10 (assuming that the interval $T$ contains the time the particle spent at $D$), whereas if the total duration is 30 seconds, then the probability is 1/30.

Namely, the dependence of the probability of arrival on the experiment duration time $T$ is a trivial fact, that immediately follows from our TOA definition.  We cannot understand why Cavendish et al. find it ``empirically implausible''. Possibly, they are using a different notion of time of arrival, although we could not think of a reasonable one where this dependence is not present. In any case, the main point of our (and others) TOA analysis is not to find the probability that the particle arrives, rather to find the probability distribution of the {\em times} at which the particle arrives. This is a conditional probability: the probability that time is $t$, given that the particle has arrived. This is the main result of \cite{qtoa}. So, we agree that the results of \cite{qtoa} may depend on the total duration $T$ of the experiment, and we carefully analyze this dependence in \cite{qtoa} and \cite{toacomparison}: e.g. the normalization issues for slow packets (as well as other regimes) have been extensively addressed in \cite{toacomparison}. In particular, Eq.~16 of \cite{toacomparison} already describes what is reported in Fig.~1 of \cite{comment}, highlighting that the slower the packet is, the faster the normalization constant diverges.

In essence, the dependence on $T$ is entirely expected with the above TOA definition, as the above simple example in the classical regime shows. It is not at all ``empirically implausible''.

\section*{``Non-arrival probability'' criticism}
The argument in the previous section also shows that Cavendish et al.'s interpretation of their un-numbered equation for the non-arrival probability ${\mathbb P}_{QC}($na$|\psi)$ is incorrect. Contrary to their claims, that equation does {\em not} describe the ``non-arrival probability''. Instead, it describes the ``probability that the particle is not found at the detector position'' (in accordance to the above definition of TOA). In the regime they consider, it is correct that such probability goes to one for $T$ going to infinity, even if the particle {\em does} arrive. As discussed above in the classical case, the probability of finding the particle at the detector is equal to the ratio between the time that the particle spends at the detector position divided by the total time of observation, assuming that $T$ contains all the period the particle spends at $D$.  If the particle crosses the detector only once and is at the detector position for 1 second, the probability of finding it there in an infinite time is $p($particle at $D)=(1 $ sec$)/(\infty$ sec$)\to 0$ (just as it is $1/10$ for an observation time of 10 seconds and $1/30$ for an observation time of 30 seconds). So, the probability of not finding it there is $p($particle not at $D)=1-p($particle at $D)\to 1$. This follows directly from our definition.

Cavendish et al.'s error is to equate the ``probability of not finding the particle at the detector position'' to the ``probability that the particle does not arrive''. They are completely different, as the above simple example shows: there the particle {\em definitely arrives} (and even spends a finite time, 1 second, at the detector), but the ``probability that it is found at the detector position'' goes asymptotically to zero for large $T$. \typeout{In other words, the ``probability that the particle does not arrive'' is equal to the
``probability that the particle was {\em never} at the detector position'', but it is {\em different} from the ``probability that the particle is not found at the detector position'' (even though it might have gone through it).}
\begin{figure}[t]
\centering
\includegraphics[width = 1 \linewidth]{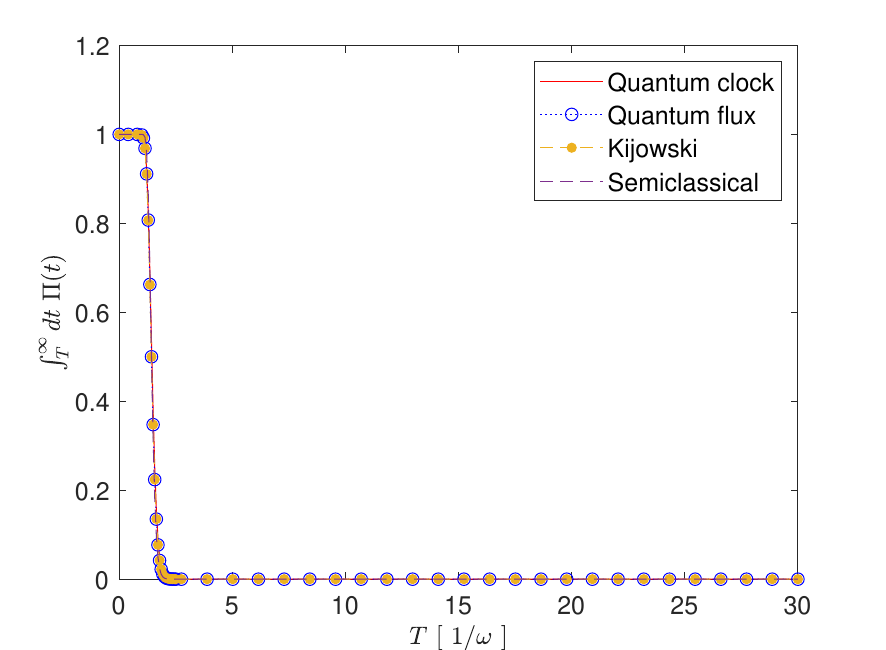}
\vspace{-.5cm}
\caption{Comparison of ${\mathbb P}_{K/F/SC}($na$|\psi)$ with $\int_{T}^\infty dt\:\Pi_{{QC}}({t})$, with $\Pi_{QC}$ normalized over a period much larger than $T$, for a detector of negligible width. The parameters of the packet are $x_0 = -10 \ l_0$, $p_0 = 7 \ \hbar/l_0$ and $\sigma_0 = 1 \ l_0$, with $l_0 = \sqrt{\hbar/m\omega}$ and $\omega$ the frequency of the harmonic trap where the particle is initially prepared. All the TOA proposals match.\label{f:cond}}
\end{figure}

Cavendish et al.~then consider the quantity ${\mathbb P}_{K/F/SC}($na$|\psi)=\int_{T}^\infty dt\:\Pi_{K/F/SC}({t})$, where $T$ is the time at which the observation is stopped. They claim this is the non-arrival probability for the Kijowski ($K$), the quantum flux ($F$), and the semi-classical ($SC$) probability distributions $\Pi({t})$. In Cavendish et al.’s Fig. 2, such quantity is compared with their un-numbered equation for the quantum clock, highlighting their differences. Two considerations: (i)~this quantity is {\em not} the probability of non-arrival even for the $K/F/SC$ proposals: by direct inspection, one can immediately see that it is the ``joint probability of finding the particle at the detector at times greater than $T$'', which is{\em , in general}, a very different situation from ``the particle not arriving during the observation time $T$''.\typeout{It matches only in very specific situations, e.g.~the one considered by Cavendish et al.: a particle initially well localized at one side of the detector and with negligible counterpropagating momentum components.} (ii)~The ``probability of not finding the particle at the detector position'', i.e.
${\mathbb P}_{QC} (na|\psi)$ in Cavendish et al.’s notation, and the ``joint probability of finding the particle at the detector at times greater than $T$'', i.e.  ${\mathbb P}_{K/F/SC} (na|\psi)$, are completely different quantities so their comparison (in Fig.~2 of Cavendish et al.) is meaningless. Instead, the calculation of the latter quantity (``joint probability of finding the particle at the
detector at times greater than $T$'') for the quantum clock, namely $\int_{T}^\infty dt\:\Pi_{QC}({t})$, gives a result that fully matches those of $K/F/SC$, as shown in Fig.~\ref{f:cond}, when the probability distribution $\Pi_{QC}({t})$ is normalized over a period much larger than $T$ (as required by the Page-Wootters mechanism to even define arrivals at times larger than $T$, as the quantity $\int_{T}^\infty dt\:\Pi_{QC}({t})$ requires). This normalization is consistent with the other TOA proposals. [Incidentally, the fact that $QC$ matches the $K/F/SC$ proposals in this regime is unwarranted: as discussed above, there is no reason why $QC$ should (or should not) agree with the others.]

In essence, Cavendish et al.'s conclusion that the ``probability of non-arrival'' is equal to the ``probability of not finding the particle at the detector'' is, quite simply, wrong, as proven above by a simple counterexample in the classical regime: a particle that spends a finite time at the detector (it has arrived) but has negligible probability of being found there for large observation times. Moreover, the differences highlighted in Cavendish et al.’s Fig.~2 are due to an incorrect comparison of different quantities.

\section*{Conclusions}
In conclusion, \cite{comment} hinges on three results: the first and second results cannot be construed as criticisms to our paper as they are direct consequences of the natural TOA definition used there (and were
already pointed out in the paper \cite{qtoa} and further analyzed in \cite{toacomparison}); the third hinges on a mistaken extrapolation by the Authors on the interpretation of the formulas of \cite{qtoa} (possibly due to a different definition of TOA they might use).

\bibliography{./refs.bib}
\end{document}